# Semantic web based Sensor Planning Services (SPS) for Sensor Web Enablement (SWE)

P.Udayakumar[1], M.Indhumathi[2]

[1] Teaching Fellow, Department of Computer Technology, MIT Campus,
Anna University Chennai, India

[2] Research Scholar, Department of Computer Science, Kajamalai Campus,
Bharathidasan University, Tiruchirappalli, India

[1]udayameister@gmail.com, [2]umindhu@gmail.com

**Abstract.** The Sensor Planning Service (SPS) is service model to define the web service interface for requesting user driven acquisitions and observation. It's defined by the Open Geospatial Consortium (OGC) Sensor Web Enablement (SWE) group to provide standardized interface for tasking sensors to allow to defining, checking, modifying and cancelling tasks of sensor and sensor data. The goal of Sensor Planning Service (SPS) of OGC – SWE is standardize the interoperability between a client and a server collection management environment. The Sensor Planning Service (SPS) is need to automate complex data flow in a large enterprises that are depend on live & stored data from sensors and multimedia equipment. The obstacle are faced in Sensor Planning Service (SPS) are (I) Observation from sensor at the right time and right place will be problem, (II) acquisition information(data) that are collected at a specific time and specific place will be problem. The above two obstacle are accomplished and obtained by the web based semantic technology in order to provide & apply the ontology based semantic rule to user driven a acquisitions and observation of Sensor Planning Service (SPS). The novelty of our approach is by adding the semantic rule to Sensor Planning Service model in SWE and we implemented Sensor Planning Service (SPS) with semantic knowledge based to achieve high standardized service model for Sensor Planning Service (SPS) of OGC – SWE.

**Keywords:** Sensor Planning Service (SPS), Open Geospatial Consortium (OGC), Sensor Web Enablement (SWE), Web Semantic technology, Ontology, acquisitions and observation.

## 1. Introduction

The recent technology in Information and communication is Wireless Sensor Network (WSN). This is the one and only technology which automate everything in this universe. Many research issues are in the WSN. The first most common research issues are how we can enable the sensor network with web. Lot of obstacle are faced when enable the sensor network with web technology. Were now Sensor Web is recent technology which will give solution to the web enabled WSN. The Open Geospatial Consortium (OGC) defines standardization for the sensor web as named Sensor Web Enablement (SWE). In this paper we focused on Sensor Planning Services (SPS) from Sensor Web Enablement (SWE) service model. First now we get into some overview about the basic things about the OGC based Sensor Web Enablement (SWE) and Wireless sensor Network (WSN).

The Wireless sensor Network is the collections of a large number of heterogeneous intelligent sensors that are spatially distributed over an environment and connected through a communication network are called distributed sensor networks (DNS). A sensor network is a computer accessible network of many, spatially distributed devices using sensors to monitor conditions at different locations, such as temperature, sound, vibration, pressure, motion or pollutants. A Sensor Web refers to web accessible sensor networks and archived sensor data that can be discovered and accessed using standard protocols and application program interfaces (APIs). In an Open Geospatial Consortium, Inc. (OGC) initiative called Sensor Web Enablement (SWE), members of the OGC are building a unique and revolutionary framework of open standards for exploiting Web-connected sensors and sensor systems of all types: flood gauges, air pollution monitors, stress gauges on bridges, mobile heart monitors, Webcams, satellite-borne earth imaging devices and countless other sensors and sensor systems. SWE presents many opportunities for adding a real-time sensor dimension to the Internet and the Web. This has extraordinary significance for science, environmental monitoring, transportation management, public safety, facility security, disaster management, utilities, Supervisory Control And Data Acquisition (SCADA) operations, industrial controls, facilities management and many other domains of activity. The OGC voluntary consensus standards setting process coupled with strong international industry and government support in domains that depend on sensors will result in SWE specifications that will quickly become established in all application areas where such standards are of use.

### 1.1 High Level Architecture of SWE

The models, encodings, and services of the SWE architecture enable implementation of interoperable and scalable service-oriented networks of heterogeneous sensor systems and client applications. In much the same way that Hyper Text Markup Language (HTML) and Hypertext Transfer Protocol (HTTP) standards enabled the exchange of any type of information on the Web, the OGC's SWE initiative is focused on developing standards to enable the discovery, exchange, and processing of sensor observations, as well as the tasking of sensor systems.
The functionality that OCG has targeted within a sensor web includes:
- Discovery of sensor systems, observations, and observation processes that meet an application's or user's immediate needs;
- Determination of a sensor's capabilities and quality of measurements;
- Access to sensor parameters that automatically allow software to process and geo-locate observations;

- Retrieval of real-time or time-series observations and coverages in standard encodings
- Tasking of sensors to acquire observations of interest;
- Subscription to and publishing of alerts to be issued by sensors or sensor services based upon certain criteria.

Within the SWE initiative, the enablement of such sensor webs and networks is being pursued through the establishment of several encodings for describing sensors and sensor observations, and through several standard interface definitions for web services.

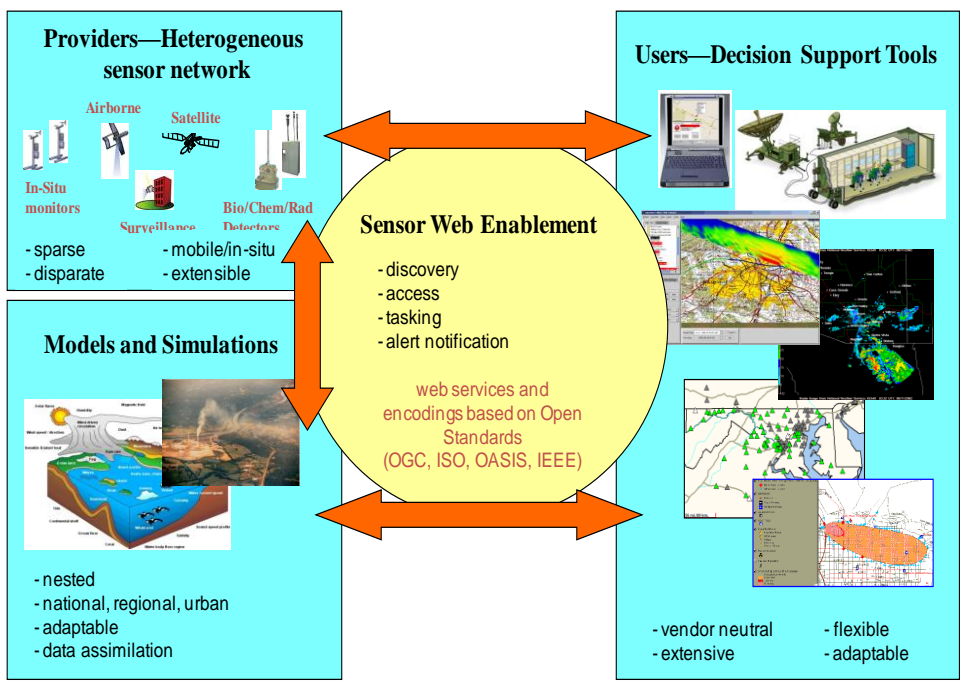

Fig 1:  The Role of Sensor Web Enablement (SWE)

Advances in digital technology are making it practical to enable virtually any type of sensor or locally networked sensor system with wired or wireless connections. Such connections support remote access to the devices' control inputs and data outputs as well as their identification and location information. For both fixed and mobile sensors, sensor location is often a vital sensor parameter. A variety of location technologies such as GPS and Cell-ID with triangulation make mobile sensing devices capable of reporting their geographic location along with their sensor collected data. When the network connection is layered with Internet and Web protocols, eXtensible Markup Language (XML) schemas can be used to publish formal descriptions of the sensor's capabilities, location, and interfaces. Then Web brokers, clients and servers can parse and interpret the XML data, enabling automated Web-based discovery of the existence of sensors and evaluation of their characteristics based on their published descriptions. The information provided also enables applications to geolocate and process sensor data without requiring *a priori* knowledge of the sensor system. Information in the XML schema about a sensor's control interface enables automated communication with the sensor system for various purposes: to determine, for example, its state and location; to issue commands to the sensor or its platform; and, to access its stored or real-time data. A Web-based application might communicate with the sensor system through a proprietary or custom interface or through an interface that implements the IEEE 1451 standard. An object-oriented approach to sensor and data description also provides a very efficient way to generate comprehensive standard-schema metadata for data produced by sensors, facilitating the discovery and interpretation of data in distributed archives.

### 1.2  Building Blocks of SWE

The *SWE architecture comprises of two major blocks:* The **information model** consists of the underlying conceptual models for encodings and the **services model** is the specification of services.

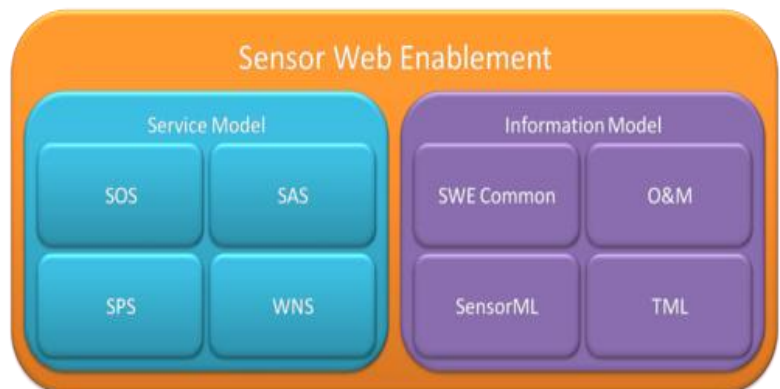

Fig 2: **Sensor web Enablement building blocks**

- Information Models
    - SWE Common
    - Observations and Measurements (O&M)
    - Sensor Model Language (SensorML)
    - TransducerML
- Services Model
    - Sensor Observation Service (SPS)
    - Sensor Planning Service (SPS)
    - Sensor Alert Service (SAS)
    - Sensor Registries.

Sensor Web Enablement standards that have been built and prototyped by members of the OGC include the following pending OpenGIS® Specifications:

1. **Observations & Measurements Schema (O&M)** – Standard models and XML Schema for encoding observations and measurements from a sensor, both archived and real-time.

2. **Sensor Model Language (SensorML)** – Standard models and XML Schema for describing sensors systems and processes; provides information needed for discovery of sensors, location of sensor observations, processing of low-level sensor observations, and listing of taskable properties.

3. **Transducer Markup Language (TransducerML or TML)** – The conceptual model and XML Schema for describing transducers and supporting real-time streaming of data to and from sensor systems.

4. **Sensor Observations Service (SPS)** - Standard web service interface for requesting, filtering, and retrieving observations and sensor system information. This is the intermediary between a client and an observation repository or near real-time sensor channel.

5. **Sensor Planning Service (SPS)** – Standard web service interface for requesting user-driven acquisitions and observations. This is the intermediary between a client and a sensor collection management environment.

6. **Sensor Alert Service (SAS)** – Standard web service interface for publishing and subscribing to alerts from sensors.

7. **Web Notification Services (WNS)** – Standard web service interface for asynchronous delivery of messages or alerts from SAS and SPS web services and other elements of service workflows.

The goal of SWE is to enable all types of Web and/or Internet-accessible sensors, instruments, and imaging devices to be accessible and, where applicable, controllable via the Web. The vision is to define and approve the standards foundation for "plug-and-play" Web-based sensor networks. Sensor location is usually a critical parameter for sensors on the Web, and OGC is the world's leading geospatial industry standards organization. Therefore, SWE specifications are being harmonized with other OGC standards for geospatial processing. The SWE standards foundation also references other relevant sensor and alerting standards such as the IEEE 1451"smart transducer" family of standards and the OASIS Common Alerting Protocol (CAP), Web Services Notification (WS-N) and Asynchronous Service Access Protocol (ASAP) specifications. OGC works with the groups responsible for these standards to harmonize them with the SWE specifications.

## 2  Background

### 2.1  Overview of Sensor Planning Services (SPS)

The Sensor Planning Service (SPS) is intended to provide a standard interface to task collection assets (i.e., satellites, other sensors, and other information gathering assets) and to the support systems that surround them. Not only will different kinds of assets with differing capabilities be supported, but also different kinds of request processing systems, which may or may not provide access to the different stages of planning, scheduling, tasking, collection, processing, archiving, and distribution of requests and the resulting observation data and information that is the result of the requests. The SPS is designed to be flexible enough to handle such a wide variety of configurations.

This standard begins with an abstract overview of the SPS interface before describing the information model for operation requests and responses in a platform-neutral manner and subsequently applying this model to a specific binding (SOAP in this case).

The Sensor Planning Services is designed and developed to enable an interoperable service by which a client can determine Collection feasibility for a desired set of collection requests for one or more sensors/platforms, or a client may submit collection requests directly to these sensors/platforms. Specifically, the document specifies interfaces for requesting information describing the capabilities of a SPS for determining the feasibility of an intended sensor planning request, for submitting such a request, for inquiring about the status of such a request, for updating or cancelling such a request, and for requesting information about further OGC Web services that provide access to the data collected by the requested task.

It defines interfaces for a service to assist in collection feasibility plans and to process collection requests for a sensor or sensor constellation. The developers and likely users of the SPS specification will be enterprises that need to automate complex information flows in large enterprises that depend on live and stored data from sensors and imaging devices. In such environments, specific information requirements give rise to frequent and varied collection requests. Quickly getting an observation from a sensor at the right time and place may be critical, and getting data that was collected at a specific place at a specific time in the past may be critical. The SPS specification specifies open interfaces for requesting information describing the capabilities of a SPS, for determining the feasibility of an intended sensor planning request, for submitting such a request, for inquiring about the status of such a request, and for updating or canceling such a request.

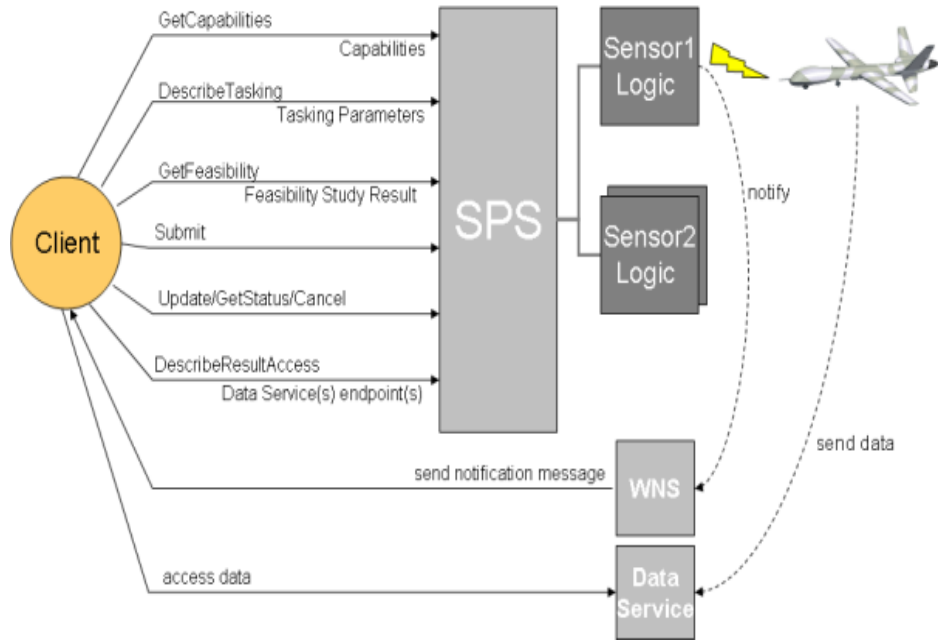

Fig 3: An exemplary workflow of an SPS controlling is shown in the following figure

**2.2   Client Server Interaction of Sensor Planning Services (SPS)**

This section explains the typical interaction between an SPS client and service. The interaction starts with the **GetCapabilities** request to explore what the service can offer. If additional information about a sensor is required, the DescribeSensor operation is used to retrieve all available information about the sensor (see Figure 4).

Next, the client needs to learn which parameters have to be set in order to task the sensor. The client sends a *DescribeTasking* request and receives a *DescribeTaskingResponse,* which defines syntax and semantic of each tasking parameter, including choices between different parameter settings, default values, and value ranges. After the client learned about the tasking parameters, it can choose to either submit a tasking request (*Submit* operation) or to perform a feasibility check (*GetFeasibility* operation).

Both operations create – if valid and accepted – a SPS assigment called task. Other operations allow to reserve and update a task, which will be discussed later on.  Requests with all required tasking parameters to the service. There is no option to use the identifier of a previous *GetFeasibility* tasking request in a subsequent *Submit/Reserve* tasking request. This lifts the burden from the service to store all *GetFeasibility* request payloads. If a task defined by the client is submitted to the service and is feasible, it is executed by the service.

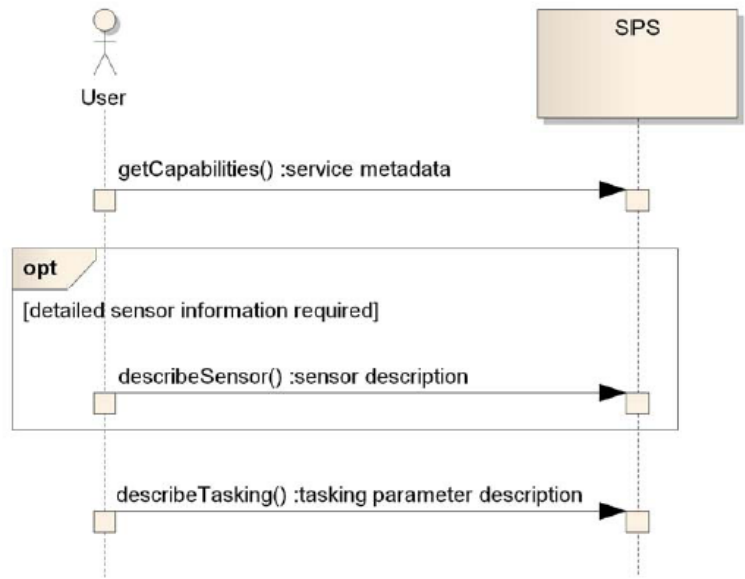

Fig 4: Client Server Interaction Part

A client may reserve a task using the *Reserve* operation. All resources required to execute the task are blocked by the service but execution does not start until the client explicitly confirms it (via the *Confirm* operation). A reservation expires at a defined point in time at which a service can reclaim all resources blocked by the reservation. Once a task is submitted/reserved, the client can *Update* or *Cancel* it. If a service cannot reserve/execute a request as provided by the client, it can provide a list of alternative parameter settings. A client can always ask for the current status of a task / tasking request via the *GetStatus* operation.

The SPS responds to *DescribeResultAccess* requests with references to all data that was produced for a given task, even if the task was cancelled or has failed. Clients can explore the references and retrieve the data gathered for this task.

The SPS service can also send notifications including *StatusReports* to inform interested clients about specific events, for example that new data has been published for a task, that a task was completed or has failed.

**2.3 Semantic Web**

The Semantic Web, as described by the W3C Semantic Web Activity, is an evolving extension of the World Wide Web in which the semantics, or meaning, of information on the Web is formally defined. Formal definitions are captured in ontologies, making it possible for machines to interpret and relate data content more effectively. The Principal technologies of the Semantic Web include the Resource Description Framework (RDF) data representation model, and the ontology representation languages RDF Schema (RDF-S) and Web Ontology Language (OWL). In addition to these representation languages, an RDF query language called SPARQL is now a W3C recommendation and the common method of querying ontological data. Many rule languages and rule engines are now capable of reasoning with Semantic Web data, including SWRL (Semantic Web Rule Language), RIF (Rule Interchange Format), and the general purpose rule engine for the Jena Semantic Web Framework.

# 3   Ontology Models for SPS

The ontology is a formal model that defines concepts and their relations in a standard language, commonly described as a "specification of a conceptualization." In practice, the Semantic Web defines several ontology languages, RDF, RDF-S, and OWL. The Resource Description Format (RDF) is a graph-based language that allows data within a domain to be linked through named relationships. An RDF graph is encoded as a set of subject-predicate-object triples which resemble the subject, verb, and object of a sentence. The subject and object are nodes in the graph and the predicate is a directional named link between the subject and object. "This simple triple structure turns out to be a natural way to describe a large majority of the data processed by machines. The subjects, verbs and objects are each identified by a Universal Resource Identifier (URI)—an address just like that used for Web pages. Thus, anyone can define a new concept, or a new verb, by defining a URI for it on the Web, RDF-S, or RDF Schema, adds the ability to define hierarchies of concepts to RDF. The Web Ontology Language (OWL) is built on top of RDF and adds a logical formalism to the language. OWL is based on a tractable subset of First Order Logic called Description Logic. The logical formalism provided by OWL, in over semantically annotated sensor observations. The ontologies dealt with in this paper are encoded in OWL.

**3.1 Sensor Planning Services Ontology**

The Sensor Planning Services (SPS) is an OGC-SWE standard which defines an XML Schema for describing planning services and its features. Within this standard, an observation (*sps:GetCapabilities*) is request to explore what the service can offer. If additional information about a sensor is required, the *DescribeSensor* operation is used to retrieve all available information about the sensor and *sps:GetFeasibility sps:GetCapabilities*,*sps:DescribeTasking*,*sps:DescribeResultAccess*, *sps:GetTask* and *sps:GetStatus* request, clients always send *Submit*/*Reserve* tasking requests with all required tasking parameters to the service. Therefore, these properties are better described as relationships of an observation. In order to encode relationships in XML, the OGC-SWE often make use of XLink, XML Linking Language, a markup language that "allows elements to be inserted into XML documents in order to create and describe links between resources.  XLink provides a framework for creating both basic unidirectional links and more complex linking structures.
It allows XML documents to:
- Assert linking relationships among more than two resources.
- Associate metadata with a link.
- Express links that reside in a location separate from the linked resources.

While XLink allows XML documents to break free of the standard tree-model and define relationships between entities, the triple-pattern approach of RDF provides a far more natural and useful approach to encoding relationships. In RDF and OWL, relationships are considered first-class objects which have many benefits over XLink, such as the ability to assign a URI to a relationship, to classify relationships into hierarchies (RDF-S and OWL), and place constraints on relationships (OWL). For these reasons, we have developed an encoding of the Observations and Measurements language in OWL. In this ontology, we have defined the previous relations, and more, in a form that may be queried and reasoned over effectively in order to derive actionable knowledge of the environment from sensor observations. (Note that the ontology captures a subset of concepts in SPS. A few notable exemptions currently include concepts related to coverage and sampling feature). The translation between O&M in OWL and SPS in XML is straightforward and thus allows Semantic web based SPS to remain SPS compliant. (From this point forward, we will refer to SPS in OWL as SPS-OWL and refer to SPS in XML as SPS-XML).

The following descriptions of relationships in SPS-OWL includes a running example of an observation from the domain of weather (concepts from eather ontology contain namespace "*w*"), encoded as a set of RDF triples. (Each line represents a triple, with the first term representing the subject, the second representing the predicate, the third representing the object, and ending with a period).

*sps:getfes_1 rdf:type sps: GetCapabilities*

## 4  Implementation of Semantic web based SPS

Here we will show some of the example SPS – xml based semantic tasking parameters representation.
SPS servers describe optional and mandatory tasking parameters. Clients use the definition to provide corresponding tasking parameter values. To ensure common understanding between client and server, a common exchange protocol is used to express both descriptions and tasking parameter values.   SPS uses the types defined in the Swe Common Data Model to define tasking parameters. The tasking parameters of a given procedure are defined in the *DescribeTaskingResponse*. Clients have to use one of the encodings provided in the contents section of the capabilities (e.g. *TextEncoding*, *XMLEncoding*, etc.) to encode the tasking parameters in the various tasking requests.

Listing 1 – example of tasking parameters corresponding to description provided by client in given encoding

```
<sps:ParameterData …>
<sps:encoding>
<swe:TextEncoding tokenSeparator="," blockSeparator="@@"/>
</sps:encoding>
<sps:values>2010-08-20T12:37:00+02:00,2010-
0820T14:30:00+02:00,Y,pointToLookAt,51.902112,8.192728,0,Y,3.5
</sps:values> </sps:ParameterData>
```

**Implementation of Channel based filtering/SPS notification topics**

When using channel based filtering, it is imperative to define which channels can be used and which notifications are sent on each channel. The OASIS WS-Topics standard defines the *TopicNamespace* type as a mean to group and describe channels/topics that belong to a specific (target) namespace. The topic namespace of this standard is defined through:

Listing 2 – SPS Topic Namespace

```
<wstop:TopicNamespace xmlns:wstop="http://docs.oasis-
open.org/wsn/t-1" xmlns:sps="http://www.opengis.net/sps/2.0"
name="SPS-Topic-Namespace"
targetNamespace="http://www.opengis.net/sps/2.0" final="true">
<wstop:Topic name="TaskEvent">
<wstop:Topic name="TaskFailure" messageTypes="sps:StatusReport"/>
<wstop:Topic name="TaskCancellation"
messageTypes="sps:StatusReport"/>
<wstop:Topic name="TaskCompletion"
messageTypes="sps:StatusReport"/>
<wstop:Topic name="TaskConfirmation"
messageTypes="sps:StatusReport"/>
<wstop:Topic name="TaskUpdate" messageTypes="sps:StatusReport"/>
<wstop:Topic name="DataPublication"
messageTypes="sps:StatusReport"/>
<wstop:Topic name="TaskReservation"
messageTypes="sps:ReservationReport"/>
<wstop:Topic name="TaskSubmission"
messageTypes="sps:StatusReport"/>
<wstop:Topic name="ReservationExpiration"
messageTypes="sps:ReservationReport"/> </wstop:Topic>
<wstop:Topic name="TaskingRequestEvent">
<wstop:Topic name="TaskingRequestExpiration"
messageTypes="sps:StatusReport"/>
<wstop:Topic name="TaskingRequestRejection"
messageTypes="sps:StatusReport"/>
<wstop:Topic name="TaskingRequestAcceptance"
messageTypes="sps:StatusReport"/>
<wstop:Topic name="TaskingRequestPending"
messageTypes="sps:StatusReport"/> </wstop:Topic>
</wstop:TopicNamespace>
```

The following table defines which events are published on which topics. In order to validate the framework discussed above, we have constructed a prototype of Semantic based SPS. Our Semantic based SPS extends the open source implementation of SPS from 52North with an ontological knowledge base in order to provide inference over sensor data and queries of high level features.

### 4.1  52North SPS

52North's SPS implementation is designed to be highly modular, and adaptable to arbitrary suitable sensor data sources, transport protocols, etc. These can be either publishers or consumers of sensor data, and may also be other web services. The Presentation Layer of 52North's

architecture defines the SPS's interface to the outside world. The default implementation has a Servlet interface that accepts requests and communicates responses via HTTP. If another transport mechanism or protocol is required, this level can be replaced without affecting the other layers of the SPS. The Visualization Layer is not part of the SPS itself, but rather corresponds to external clients that interact with the SPS. These can be either publishers or consumers of sensor data, and may also be other web services. The Presentation Layer of 52North's architecture defines the SPS's interface to the outside world. The default implementation has a Servlet interface that accepts requests and communicates responses via HTTP. If another transport mechanism or protocol is required, this level can be replaced without affecting the other layers of the SPS. The next level is the Business Layer, which receives requests from the Presentation Layer, handles them as appropriate, and returns a response. The Business Layer contains the logic for decoding requests and encoding responses.

The main entry-point from the Presentation Layer is the *RequestOperator* object, which validates incoming requests, determines the type of request, and dispatches accordingly. Each operation supported by the SPS(*sps:GetFeasibilitysps:GetCapabilities*,*sps:DescribeTasking*,*sps:DescribeResultAccess*, *sps:GetTask* and *sps:GetStatus*, etc.) is embodied by a Listener object which handles the corresponding incoming request.

The Listener objects may be configured externally during deployment of the service. The individual Listeners handle high-level translation of the request into an internal format which is then used to query the respective object in the Data Layer and compose the response. The final layer of the 52North architecture is the Data Layer. The Data Layer is an abstraction of a sensor data source through Data Access Objects (DAO). Each DAO represents a particular interface to the sensor data from the point of view of one of the SPS's operations. For each Listener object in the Business Logic Layer, there is a corresponding DAO object in the Data Layer. The DAO objects are used by their respective Listener objects to obtain the data pertaining to a query. The abstraction provided by the DAOs and the Data Layer is what allows the 52North's SPS implementation to be so easily adapted to new sources of sensor data. For each operation that must be supported, all that is required is a new DAO that works with the data source. The default implementation shipped with 52North uses a PostGIS database with a custom database schema to store observation data, while sensor descriptions are stored on the file system in XML files (using SensorML or TransducerML).

## 5   Conclusion and Future work

A synthesis of the Sensor Web Enablement standards defined by the OGC and the Semantic Web languages defined by the W3C provides a platform for integration and reasoning over sensor observations in order to attain shared knowledge of an environment. This platform is broadly termed the Semantic Sensor Web, of which Semantic based SPS is a principal component. In the preceding sections we have described how this is accomplished by modeling the domain of sensors and sensor observations in a suite of ontologies, adding semantic annotations to the sensor data, using the ontology models to reason over sensor observations, and extending an open source SOS implementation with our semantic knowledge base. In the future, we hope to incorporate an addictive reasoning engine as well as expand the Semantic Sensor Web platform. Addictive reasoning is often described as *inference to the best explanation*. In the sensors domain, a phenomenon is an effect that could have been caused (or could be explained) by many possible features, or real-world objects and events. An addictive reasoning engine would provide the ability to reason from sensor observations of phenomena to possible hypothesis, or possible features, of the environment. Through an implementation of the Semantic based SPS transactional profile (*RegisterSensor*, *GetCapabilities*), and translation from SPS-XML to SPS-OWL, standard implementations of SOS may take advantage of the addictive reasoning capabilities of Semantic based SPS in a modular, distributed, and standards-based environment. In addition, we are planning on extending the Semantic Sensor Web platform beyond SPS-OWL. Such plans include developing an OWL version of Sensor Model Language (SML-OWL) and Sensor Alert Service (Semantic based SAS).It is our belief that the addition of semantics to the OGC Sensor Web Enablement standards provides an improved platform for discovering, accessing, controlling, and reasoning over sensors and sensor observation data on the Web.